\documentclass[fleqn,usenatbib,useAMS,preprint,usedcolumn,onecolumn]{mnras}

\usepackage{newtxtext,newtxmath}

\usepackage[T1]{fontenc}
\usepackage{ae,aecompl}

\DeclareRobustCommand{\VAN}[3]{#2}
\let\VANthebibliography\thebibliography
\def\thebibliography{\DeclareRobustCommand{\VAN}[3]{##3}\VANthebibliography}

\usepackage{graphicx}	
\usepackage{amsmath}	
 
\usepackage{amssymb}	
\usepackage{amsfonts}
\usepackage{multicol}        
\usepackage{bm}         
\usepackage{pdflscape}	
\usepackage{lineno}
\usepackage{hyperref}
\usepackage[pdftex]{color}
\usepackage{xcolor}

\title[Markov analysis]{Statistical analysis of stochastic magnetic fluctuations in space plasma\\ 
based on the \textit{MMS} mission}

\author[W. M. Macek and D. W\'{o}jcik]{
Wies{\l}aw M. Macek,$^{1,2}$\thanks{E-mail: macek@uksw.edu.pl, macek@cbk.waw.pl  (WMM)}
and Dariusz W\'{o}jcik$^{1,2}$\thanks{E-mail: d.wojcik@uksw.edu.pl, dwojcik@cbk.waw.pl  (DW)}
\\
$^{1}$Institute of Physical Sciences, Faculty of Mathematics and Natural Sciences, 
Cardinal Stefan Wyszy\'{n}ski University, W\'{o}ycickiego 1/3, 01-938 Warsaw, Poland\\
$^{2}$Space Research Centre, Polish Academy of Sciences, Bartycka 18A, 00-716 Warsaw, Poland
}

\date{Accepted 2023 August 20. Received 2023 August 12; in original form 2023 May 17}

\pubyear{2023}

\begin{document}
\label{firstpage}
\pagerange{\pageref{firstpage}--\pageref{lastpage}}
\maketitle

\begin{abstract}
Based on the \textit{Magnetospheric Multiscale} (\textit{MMS}) mission
we look at magnetic field fluctuations in the Earth's magnetosheath. 
We apply the statistical analysis using a Fokker--Planck equation 
to investigate processes responsible for 
stochastic fluctuations in space plasmas. 
As already known, turbulence in the inertial range of hydromagnetic scales exhibits Markovian features.
We have extended the statistical approach to much smaller scales in space, 
where kinetic theory should be applied. 
Here we study in detail and compare the characteristics of 
magnetic fluctuations behind the bow shock,
inside the magnetosheath, and near the magneto\-pause. 
It appears that the first Kramers--Moyal coefficient 
is linear and the second term is quadratic function of magnetic increments,
which describe drift and diffusion, correspondingly, in the entire magnetosheath. 
This should correspond to a generalization of Ornstein--Uhlenbeck process.
We demonstrate that the second order approximation of the Fokker--Planck equation 
leads to non-Gaussian kappa distributions of the probability density functions.  
In all cases in the magnetosheath, the approximate power-law distributions are recovered. 
For some moderate scales we have the kappa distributions 
described by various peaked shapes with heavy tails. 
In particular, for large values of the kappa parameter 
this shape is reduced to the normal Gaussian distribution. 
It is worth noting that for smaller kinetic scales 
the rescaled distributions exhibit a universal global \emph{scale\discretionary{-}{-}{-}invariance},
consistently with the stationary solution of the Fokker--Planck equation. 
These results, especially on kinetic scales, could be important 
for a better understanding of the physical mechanism 
governing turbulent systems in  space and astrophysical plasmas.
\end{abstract}




\section{Introduction}  
\label{sec:wm:int} 

Turbulence is a complex phenomenon that notwithstanding progress in (magneto-)hydrodynamic simulations
is still a challenge for natural sciences  \citep{Fri95}, 
and physical mechanisms responsible for turbulence cascade are  not clear \citep{Bis03}. 
Fortunately, collisionless solar wind  plasmas can be considered natural laboratories 
for investigating this complex dynamical system  \citep{BruCar16}.  
Fluctuations of magnetic fields play an important role in space plasmas, 
leading also to a phenomenon known as magnetic reconnection \citep[e.g.,][]{Bur95,Tre09}.
  
Turbulent magnetic reconnection is a process in which energy can proficiently be shifted 
from a magnetic field to the motion of charged particles.
Therefore, this process responsible the redistribution of kinetic and magnetic energy in space plasmas 
is pivotal to the Sun, Earth, as well as to other planets and generally across the whole Universe. 
Reconnection also impedes the effectiveness of fusion reactors and regulates geospace weather 
which can affect contemporary technology such as the Global Positioning System (GPS) navigation, 
modern mobile phone networks, including  electrical power grids.
Electric fields  responsible for reconnection in the Earth's magnetosphere
has been observed on kinetic scales by the \textit{Magnetospheric Multiscale} (\textit{MMS}) mission
\citep{Macet19a,Macet19b}. 
Certainly, reconnection in the magnetosphere is related to turbulence at small scales \citep{Karet14}. 

Basically, in a Markov process, given an initial probability distribution function (PDF), 
the transition to the next stage can fully be determined. 
It is also interesting here that we can prove and demonstrate 
the existence of such a Markov process experimentally. 
Namely, without relying on any assumptions or models for the underlying stochastic process
we are able to extract the differential equation of this Markov 
process directly from the collected experimental data. 
Hence this Markov approach appears to be a bridge between 
the statistical and dynamical analysis of complex physical systems.
There is a substantial evidence based on statistical analysis that 
stochastic fluctuations exhibits Markov properties \citep{PedNov94,Renet01}.
We have already proved that turbulence 
has Markovian features 
in the inertial range of hydromagnetic scales  \citep{StrMac08a,StrMac08b}.
Admittedly, for turbulence inside the inertial region of magnetized plasma, 
the characteristic spectra should be close to  standard \citeauthor{Kol41} (\citeyear{Kol41})  
power-law type with exponent: $-5/3 \approx -1.67$ \citep{Kol41}
and \citeauthor{Kra65} (\citeyear{Kra65}) 
power-law spectrum with exponent: $-3/2$, 
but surprisingly, the absence of these classical spectra, 
especially on smaller scales, seems to be the rule. 

Moreover, we have also confirmed clear breakpoints in the magnetic energy spectra in the Earth's magnetosheath,
which occur near the ion gyrofrequencies just behind the bow shock, inside the magnetosheath,
and before leaving the magnetosheath. 
Namely, we have observed that the spectrum steepens at these points to power exponents 
in the kinetic range from  -5/2 to -11/2 for the magnetic field data of the highest resolution 
available within the \textit{MMS} mission \citep{Macet18}.
Therefore, we would like to investigate the Markov property of 
stochastic fluctuations outside this inertial region of magnetized plasma on small scales, 
when the slopes are consistent with kinetic theory.
 
It should also be noted that based on the measurements of magnetic field fluctuations 
in the Earth's magnetosheath gathered onboard the \textit{MMS} mission,
we have recently extended these statistical results to  much smaller scales, 
where kinetic theory should be applied \citep{Macet23}.
Here we compare the characteristics of 
stochastic fluctuations behind the bow shock,
inside the magnetosheath, and near the magneto\-pause. 
In this paper, we therefore present the results of our comparative analysis, 
where we check whether the solutions of the Fokker--Planck (FP) equation are consistent  
with experimental PDFs in various regions of the magnetosheath.

In Section~\ref{sec:wm:d}, a concise description of the \textit{MMS} mission and the analyzed data is provided, 
while the Section~\ref{sec:wm:meth} outlines used stochastic mathematical and statistical methods.
The vital results of our analysis are presented in Section~\ref{sec:wm:res}, 
which demonstrates that the solutions of the FP  
equation are in good agreement with empirical PDFs. 
Finally, Section~\ref{sec:wm:con} emphasizes the significance of stochastic Markov processes 
in relation to turbulence in space plasmas, 
which exhibit a universal global \textit{scale-invariance} across the 
kinetic domain.

\section{Data}
\label{sec:wm:d}

The \textit{MMS} mission, which begun in March 12, 2015 and is still in operation, 
delves into the connection and disconnection of the Sun's and Earth's magnetic fields. 
Four spacecraft (namely \textit{MMS 1} -- \textit{MMS 4}), 
which carry identical instrument suites, are orbiting near the equator to observe magnetic turbulence  in progress. 
They are formed into a pyramid-like formation. 
Each satellite has an octagonal form that is around 3.5 meters in breadth and 1.2 meters in height. 
The satellites rotate at Three  Revolutions Per Minute during scientific operations. 
Position data is supplied by ultra-precise GPS apparatus, 
while attitude is sustained by four stellar trackers, two accelerometers, and two solar sensors. 
All of the spacecraft have identical instruments to measure plasmas, magnetic and electric fields, 
and other particles with remarkably high (milliseconds) time resolution and accuracy. 
This allows us to figure out if reconnection takes place in an individual area, 
everywhere within a broader area simultaneously, or traversing through space. 
The \textit{MMS} studies the reconnection of the solar and terrestrial magnetic fields 
in both the day and night sides of Earth's magnetosphere, 
which is the only place where it can be directly observed by spacecraft.
In our previous studies we have observed reconnection in the Earth's magnetosphere
involving small kinetic scales \citep{Macet19a,Macet19b}.

We have further examined fluctuations of all components of the magnetic fields  
$\mathbf{B} = (B_x, B_y, B_z)$ in the Geocentric Solar Ecliptic (GSE) coordinates,
with the magnitude strength $B = |\mathbf B|$
(square root of the sum of the squares of the field components), 
which have been taken from the \textit{MMS} Satellite No.~1 (\textit{MMS 1}),
located just beyond the Earth's bow shock (BS).
In this way, we have shown that magnetic 
fluctuations exhibits Markov character also on very small kinetic scales \citep{Macet23}. 
Moreover, we have noticed  that in all components the Markovian features are quite  similar. 
Here, we would like to further investigate statistical properties of magnetic 
fluctuations in various regions of the magnetosheath.  
The spacecraft trajectories in the magnetosheath, in three different regions, namely:
\begin{itemize}
    \item (a) just behind the bow shock (BS),
    \item (b) deep inside the magnetosheath (SH), and 
    \item (c) near the magnetopause (MP),
\end{itemize}
which have been shown in Fig.~1 of \citep[]{Macet18}. 

Therefore, we would like to look at the measurements of the magnetic field strength $B = |\mathbf{B}|$,
but now at various regions of the magnetosheath. 
To investigate magnetosheath 
stochastic fluctuations, now we have chosen the same three different time intervals samples, 
which correspond to Table~1 (List of Selected \textit{MMS 1} Interval Samples) of Ref. \citep{Macet18}. 
In cases (a) and (c) of approximately 5 minutes and 1.8 minutes respective intervals, 
we use burst type observations from the FGM (FluxGate Magnetometer) sensor  
with the highest resolution ($\Delta t_B$) of 7.8 ms (128 samples per second) 
with 37,856 and 13,959 data points, correspondingly. 
However, in the other case (b), between the bow shock and the magnetopause, 
where only substantially lower resolution 62.5--125 ms in survey  mode 
(8--16 samples per second) data are available, 
we have a much longer interval lasting 3.5 h with 198,717 data points with $\Delta t_B = 62.5$ s. 
All of the data are publicly available on the website: \url{http://cdaweb.gsfc.nasa.gov}, 
which is hosted by NASA.

\begin{figure} 
\centering
\includegraphics[scale=0.6]{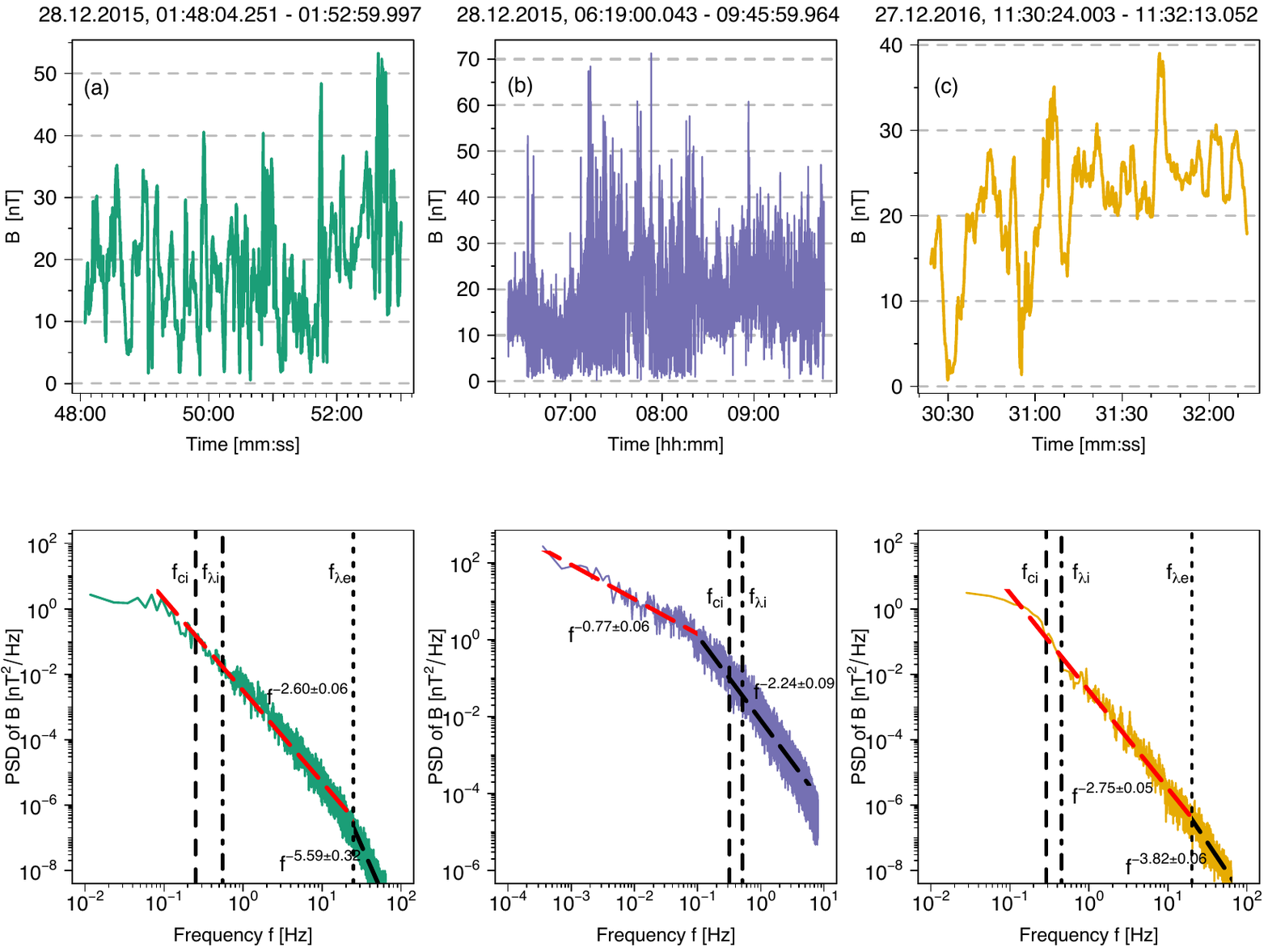}
\caption{Time series of the magnetic field strength  $B = |\bold B|$ 
of the \textit{MMS} data with the corresponding spectra 
in the magnetosheath (a) near the bow shock (BS), (b) inside the magnetosheath (SH), 
and (c) near the magnetopause (MP)
plotted with three different colors. 
Average ion gyrofrequency ($f_{ci}$), 
as well as a characteristic Taylor's shifted frequencies for ions ($f_{\lambda i}$) and electrons ($f_{\lambda e}$) 
are shown by the dashed, dashed-dotted, and dotted lines, respectively,
see Table~1 of Ref.~(Macek et al. 2018).
}
\label{f:wm:data}
\end{figure}

Admittedly, the gaps in time series, which commonly appear in the data gathered from space missions,   
can have a considerable impact on the conclusions that can be derived from statistical 
analysis based on experimental data. 
One of the powerful but simple tools used to cope with this problem 
is a linear interpolation method between points, 
which we have used, if necessary, to fill these gaps in the analyzed data sets.
Therefore, in Fig.~\ref{f:wm:data} on the upper side of each case (a) -- (c) from left to right, 
we have depicted time series of the magnetic field strength $B = |\mathbf{B}|$. 
Whereas on the bottom side of each case, we have shown  
the respective power spectral density (PSD) 
obtained using the method proposed by \cite{Wel67}.

The calculated average ion and electron gyrofrequencies are as follows: 
in case (a) $f_{ci} = 0.25$ Hz, and $f_{ce} = 528$ Hz; 
case (b) $f_{ci} = 0.24$ Hz and $f_{ce} = 510$ Hz; 
case (c) $f_{ci} = 0.29$ Hz and $f_{ce} = 609$ Hz \citep{Macet18}. 
In addition, employing the hypothesis according to \cite{Tay38},
relating time and space scales
in this way: $l = \varv_\mathrm{sw} \cdot \tau$, where $l$ is a spatial scale 
and $\varv_\mathrm{sw}$ is the mean velocity of the solar wind flow in the magnetosheath, 
we estimate characteristic inertial frequencies for ions and electrons: 
in case (a) $f_{\lambda i} = 0.55$ Hz and $f_{\lambda e} = 24.5$ Hz; 
case (b) $f_{\lambda i} = 0.41$ Hz and $f_{\lambda e} = 18.1$ Hz; 
case (c) $f_{\lambda i} = 0.45$ Hz and $f_{\lambda e} = 20.1$ Hz. 
We have marked these values on each graph of power spectral density. 
In case (a) the obtained spectral exponent is about $-2.60 \pm 0.06$ somewhat 
steeper, before the $f_{\lambda e} = 24.5$ Hz threshold and undoubtedly more steepen 
than the \citeauthor{Kol41} (\citeyear{Kol41}) (-5/3) or
\citeauthor{Kra65} (\citeyear{Kra65}) (-3/2) slopes. 

On the other hand, 
outside the inertial range of scales large  spectral exponents has been reported 
from the \textit{Cluster} multi-spacecraft mission \citep{Sahet09},
the \textit{WIND} data \citep{Bruet14},
including the proposed explanation of nature of solar wind magnetic fluctuations on kinetic scales 
based on the missions \citep[e.g.,][]{Lioet16,Robet16}.
Owing to unprecedented high 7.8 ms  
time resolution of magnetometer data in the \textit{MMS} mission available in burst mode,
we also see that in case (a) the slope is of $-2.60\pm0.06$ (close to -5/2) above $f_{\lambda e} = 24.5$ Hz.
This is further followed by an even steeper spectrum 
with the slope of $-5.59\pm0.32$ (close to -11/2 or -16/3).
Because of a substantially lower survey data resolution of 62.5 ms 
in case (b) the spectrum  with $-2.24\pm0.09$ ($\approx -7/3$) 
is steeper than the \citeauthor{Kol41} (\citeyear{Kol41}) ($-5/3$) spectrum 
only after the visible breakpoint in the slope,
which lies at $f = 0.12$ Hz, i.e. near the ion gyrofrequency $f_{ci} = 0.24$ Hz, 
while more gentle slope of $-0.77\pm0.06$ is observed before this breakpoint. 
Finally, in case (c), similarly as in case (a) using burst  data, 
the spectral exponent of $-2.75\pm0.05$ is again steeper before, 
and even more with the exponent $-3.82\pm0.06$ (close to -7/2)  
after the observed breakpoint that lies at around the electron 
\citeauthor{Tay38}'s (\citeyear{Tay38}) 
shifted frequency $f_{\lambda e} = 20$~Hz, 
as discussed by \cite{Macet18}.
This shows that the observed stochastic nature of fluctuations in the sub-ion scale 
could be due to the interaction between coherent structures \citep{Perrone16,Perrone17},
and a very high slope of -16/3 is possibly related to 
the dissipation of the kinetic Alfv\'{e}n waves \citep[e.g.,][]{Schet09}.

\section{Methods of Data Analysis}
\label{sec:wm:meth}

As usual, we use  the fluctuations of the magnetic fields $B = |\mathbf{B}|$, 
which describe this turbulent system at each time $t > 0$.
Therefore, with a given time scale $\tau_i > 0 \; \; \forall_i$, 
one can typically define the increments of this quantity as follows:
\begin{equation}
b_{i} (t) := B (t+\tau_i) - B (t),
\label{e:wm:tau}
\end{equation}
and, assuming an arbitrary $\tau_i > 0$, 
it can be labeled as $b_{\tau}$ or $b$ for simplicity in the following sections. 

We assume that the fluctuations of increment $b_{\tau}$ in a larger time scale $\tau$ 
are transferred to smaller and smaller scales. 
Therefore, 
stochastic fluctuations may be regarded as a stochastic process in scale  
with the $N$-point joint (transition) conditional probability density function 
denoted by $P(b_1,\tau_1 | b_2,\tau_2, \ldots, b_N,\tau_N)$. 
In this case, the conditional probability density function is defined by default as:
\begin{equation}
P(b_i,\tau_i|b_j,\tau_j) = \frac{P(b_i, \tau_i; b_j, \tau_j)}{P(b_j,\tau_j)},
\label{e:wm:cond}
\end{equation}
with the  marginal (unconditional) probability density function, $P(b_j,\tau_j)$, 
and the joint probability function, $P(b_i, \tau_i; b_j, \tau_j)$, 
of finding the fluctuations $b_i$ at a scale $\tau_i$ and $b_j$ at a scale $\tau_j$, for $0 < \tau_i < \tau_j$. 
In the same way, we may construct the conditional probability densities 
for any longer sequences of increments $b$.

The stochastic process is Markovian if the conditional probability function depends 
only on the initial values $b_1$ and $b_2$ at the time scales $\tau_1$ and  $\tau_2$, 
but not on $b_3$ at the next larger scale  $\tau_3$, and so on, i.e., for any $i = 1, \ldots, N$ we have:
\begin{equation} 
\label{e:wm:mar}
P(b_1,\tau_1|b_2,\tau_2) = P(b_1,\tau_1|b_2,\tau_2 , \ldots , b_N,\tau_N),
\end{equation}
for $0 < \tau_1 < \tau_2 < \ldots < \tau_N$. 
Basically, the Markov process can be determined by 
the initial conditional probability function $P(b_1,\tau_1|b_2,\tau_2)$. 
Strictly speaking, the future states of the process are conditionally independent of past states.
Because of this relation, the conditional probabilities are also called transition probabilities, 
while the property of Eq.~(\ref{e:wm:mar}) is known as a \textit{memorylessness}.

One of the generalizations of Eq.~(\ref{e:wm:mar}) is called the Chapman--Kolmogorov (CK) condition, 
which is given by the equation \citep{Ris96}:
\begin{equation} 
\label{e:wm:chk}
P(b_1,\tau_1|b_2,\tau_2) = \int_{-\infty}^{+\infty} P(b_1,\tau_1|b',\tau') P(b',\tau'|b_2,\tau_2) d b', 
\end{equation}
for $\tau_1 < \tau' < \tau_2$. 
This equation can be interpreted in the following way: 
the transition probability from $b_2$ at a~time scale $\tau_2$ to $b_1$ 
at a~time scale $\tau_1$ is the same as a product of the transition probability from $b_2$ 
at a~time scale $\tau_2$ to $b'$ at a~time scale $\tau'$, 
and the transition probability from $b'$ at a~time scale $\tau'$ to $b_1$ at a~time scale $\tau_1$, 
for all possible $b'$'s. 
Let us emphasize here, that such a generalization is a necessary condition 
for a stochastic process to be the Markov process.

Next, from the CK condition of Eq.~(\ref{e:wm:chk}), by using a standard  
series expansion, one can derive a corresponding Kramers--Moyal backward expansion 
with an infinite number of terms.
Backward expansions are equations of evolution of probability $P(b,\tau|b', \tau')$,  
where we differentiate with respect to $b$. 
This equation has the following differential form \citep[Section~4.2]{Ris96}:
\begin{equation} \label{e:wm:fpe}
    -\frac{\partial}{\partial \tau} P(b,\tau|b', \tau') = \sum_{k = 1}^{\infty}  \left(-\frac{\partial}{\partial b}\right)^{k} D^{(k)}(b,\tau) P(b, \tau|b',\tau'), 
\end{equation}
where it is important to note that the differential symbol acts 
on both $D^{(k)}(b,\tau)$ and $P(b,\tau|b', \tau')$ coefficients. 
Since the solutions of the forward and backward KM equations are equivalent, 
then without loss of generality, we can label it as KM expansion. 
Formally, $D^{(k)}(b,\tau)$ are called KM coefficients,
which in this way are defined as the limit at $\tau \to \tau'$ 
of $k$-th power of conditional moments \citep[see][]{Ris96}: 
\begin{align}
\noindent 
D^{(k)}(b,\tau) & = \frac{1}{k!} \lim_{\tau \rightarrow \tau'} \frac{1}{\tau - \tau'} M^{(k)} (b, \tau, \tau'), 
\label{e:wm:Dk} \\ 
M^{(k)} (b, \tau, \tau') & = \int_{-\infty}^{+\infty} (b' - b)^{k}  P(b',\tau'|b,\tau) d b'. 
\label{e:wm:Mk} 
\end{align}
Ideally, using the conditional moments $M^{(k)}(b, \tau, \tau')$, 
the KM coefficients can be evaluated, 
though they cannot be obtained directly from the analyzed data. 
While these conditional moments can be calculated from the empirical observations, 
the $D^{(k)}(b, \tau)$ coefficients can only be obtained by extrapolation in the limit $\tau \to \tau'$ 
according to Eqs.~(\ref{e:wm:Dk}) and (\ref{e:wm:Mk}), 
but these formulae can not be applied explicitly. 

One of the popular extrapolation methods for this problem is a~use of piecewise linear regression model with breakpoints. 
This is a type of regression models, which allows multiple linear models to fit to the analyzed data. 
The crucial objective of this method is an accurate estimation of a number of breakpoints.  
First, in order to estimate the best breakpoint position, we have evaluated every value within a specified interval 
and looked at the value of logarithmic transformation of the likelihood function 
(also known as \textit{log-likelihood} function) of each adjusted model. 
Naturally, the highest value of this function provides the optimal breakpoint. 
Further, to select (and estimate) the best possible number of breakpoints of the segmented relationship, 
we have used the standard \citeauthor{AIC73} (\citeyear{AIC73}) Information Criterion ($\mathrm{AIC}$) 
and Bayesian Information Criterion ($\mathrm{BIC}$) \citep{BIC78}.
Nonetheless, the truly similar results are obtained when the lowest time resolution is taken. 
Thus, in our case, we have a simple approximation of the KM coefficients, 
which is given by:
\begin{equation} 
    \label{e:wm:cef_app}
    D^{(k)}(b,\tau) = \frac{1}{k!} \frac{1}{\Delta t} M^{(k)} (b, \tau, \tau'),
\end{equation}
where a $\Delta t$ is a given lowest time resolution of the time series. 
It is also interesting to note that $D^{(k)} (b,\tau)$ coefficients 
show the same dependence on $b$ as $M^{(k)} (b, \tau, \tau')$.
This simplification substantially decrease the time required 
to obtain the results numerically.

Now, in order to find the solution of Eq.~(\ref{e:wm:fpe}), 
it is necessary to determine the number of terms of the right hand side (RHS) of this equation that need to be considered.
According to Pawula's theorem, 
the KM expansion of a positive transition probability $P(b, \tau|b', \tau')$ 
may end after the first or second term \citep[e.g.,][Section~4.3]{Ris96}. 
If it does not end after the second term, 
then the expansion must contain an infinite number of terms. 
On the other hand, if the second term is the last  one, namely $D^{(k)}(b,\tau) =  0$ for $k \ge 3$, 
then the KM expansion of Eq.~(\ref{e:wm:fpe}) leads to the following particular 
formula:
\begin{equation} 
\label{e:wm:fop}
-\frac{\partial}{\partial \tau} P(b,\tau|b',\tau') =  \Bigg[-\frac{\partial}{\partial b} D^{(1)}(b,\tau) + \frac{\partial^{2}}{\partial b^{2}} D^{(2)}(b,\tau)\Bigg] P(b,\tau|b',\tau'),
\end{equation} 
with the well-known FP operator {$\mathcal{L}_\mathrm{FP}(b, \tau)$} in the squared parenthesis \citep[e.g.,][Eqs.~5.1~and~5.2]{Ris96}
governing the evolution of the probability density function $P(b,\tau|b',\tau')$ and is called the 
FP equation (also known as a forward Kolmogorov equation). 
It has been primarily used for the Brownian motion of particles,
but now Eq.~(\ref{e:wm:fop}) defines a generalized Ornstein--Uhlenbeck process. 
Strictly speaking, this is a linear second-order partial differential equation of a parabolic type. 
By solving the FP equation, it is possible to find distribution functions 
from which any averages (expected values) of macroscopic variables can be determined by integration. 
If the relevant time-dependent solution is provided, 
this equation can be used to not only describe stationary features, 
but also the dynamics of systems.

The first term $D^{(1)}(b, \tau)$ and a second term $D^{(2)}(b, \tau) > 0$ 
determining the FP Equation~(\ref{e:wm:fop}) 
are responsible for the drift and diffusion processes, respectively. 
The former process accounts for the deterministic evolution of the stochastic process (as a function of $b$ and $\tau$). 
The latter process modulates the amplitude of the $\delta$-correlated Gaussian noise $\Gamma(\tau)$ 
(which is known as the Langevin force -- the fluctuating force $F_f(\tau)$ per unit mass $m$), 
that fulfills the normalization conditions: 
$\langle \Gamma(\tau) \Gamma(\tau') \rangle = 2 \delta (\tau - \tau')$, 
where $\delta$ is a Dirac delta function
and $\langle \Gamma(\tau) \rangle = 0$ \citep[see][]{Ris96}. 
Thus, in the equivalent approach another complementary equation arises:
\begin{equation} 
     \label{e:wm:lan}
    -\frac{\partial b} {\partial \tau} = D^{(1)}(b,\tau) + \sqrt{D^{(2)}(b,\tau)} \cdot \Gamma (\tau),
\end{equation}
which is formally called the Langevin equation. 
Here we have used  the \citeauthor{Ito44} (\citeyear{Ito44})  
definition, that is missing a spurious drift \citep[e.g.,][Section~3.3.3]{Ris96}, 
hence the drift coefficient $D^{(1)}$ occurs directly, 
unlike in the \citeauthor{Str68} (\citeyear{Str68})  
definition. 
Admittedly, the \citeauthor{Ito44} (\citeyear{Ito44})  
definition is more difficult to interpret and analyze, 
because of the new rules for integration and differentiation that must be used. 
Although, owing to a powerful apparatus, which is the It\^{o} Lemma, 
it allows us to deal with stochastic processes analytically.
Anyway, here again, all higher KM coefficients $D^{(k)}$ for $k \ge 3$ are equal to zero. 
Note that the negative signs on the left hand side (LHS) of Eqs.~(\ref{e:wm:fop}) and~(\ref{e:wm:lan}) 
show that the corresponding transitions proceed backward to smaller and smaller scales.

Next, because the differentiating in the FP  operator in Eq.~(\ref{e:wm:fop})
should act on both the KM coefficients and the conditional probability density $P(b,\tau|b',\tau')$ by performing 
relatively simple transformations,  
it can be rewritten in the following expanded form \citep[Eq.~45.3a]{Ris96}:
\begin{eqnarray} 
     \label{e:wm:fop2}
    -\frac{\partial}{\partial \tau} P(b,\tau|b', \tau') =  D^{(2)}(b,\tau|b', \tau') \frac{\partial^2}{\partial b^2} P(b,\tau|b', \tau') + & {} & \nonumber\\
 +  \Big[2 \frac{\partial}{\partial b} D^{(2)}(b,\tau|b', \tau') - D^{(1)}(b,\tau|b', \tau')\Big] \frac{\partial}{\partial b} P(b,\tau|b', \tau') {} \nonumber\\ 
 +  \Big[\frac{\partial^2}{\partial b^2} D^{(2)}(b,\tau|b', \tau') - \frac{\partial}{\partial b} D^{(1)}(b,\tau)\Big] P(b,\tau|b', \tau').
\end{eqnarray}
Formally, Eq.~({\ref{e:wm:fop2}) resulting from the FP Equation~(\ref{e:wm:fop}) 
is the second-order parabolic partial differential equation.

It is also worth mentioning that this equation is the generalization of the case of thermal conductivity diffusion equation, 
which can be solved with the initial and boundary conditions 
$P(b,\tau=0|b', \tau'=0) = p(b,b')$ and  $P(b=0,\tau|b'=0, \tau') = 0$, 
respectively, using the method of separation of variables. 
The solution of nonstationary FP Equation~(\ref{e:wm:fop2}) can well be approximated numerically, i.e. by the difference method.
The master curve for the probability density function $P(b,\tau)$ of Eq.~(\ref{e:wm:fop2}) 
can readily be evaluated by the stationary solution $p_s(b, \tau)$ of Eq.~(\ref{e:wm:fop}), 
which is given by
\begin{equation} \label{e:wm:sta}
    \frac{\partial}{\partial b} \Big[ D^{(2)}(b,\tau) p_s(b, \tau) \Big] = D^{(1)}(b,\tau)  p_s(b, \tau)
\end{equation}
that results from comparing the LHS of Eq.~(\ref{e:wm:fop}) with zero.

\section{Results} 
\label{sec:wm:res}

In order to inspect processes responsible for 
stochastic fluctuations in space plasma, 
we have applied the methods described in Section~\ref{sec:wm:meth} 
to small-scale in cases (a) and (c) and medium-scale in case (b) 
fluctuations of the magnetic field $B = |\mathbf{B}|$ in the Earth's magnetosheath. 
In general, the approach presented in this paper could be  applied under a few important conditions 
that should be tested as preliminary procedures \cite[see][]{Rinet16}. 
The first condition is that time series data must be stationary. 
If they were non-stationary, then the conditional moments 
given by Eq.~(\ref{e:wm:Mk}) are not essentially meaningful. 
The second condition is that the process should be Markovian, 
i.e., the present state should only depend on the preceding state. 
The third condition is that the Pawula's theorem must hold, 
as discussed in Section~\ref{sec:wm:meth}.

Having this in mind, we have started with the brief analysis and description of the relevant time series 
and the corresponding graphs of power spectral densities (PSD). 
Next, we have checked 
stationarity of all analyzed time series \cite[see, e.g.,][]{Mac98}. 
To show that a Markov processes approach is suitable in our situation,
we have moved forward to the verification of the necessary CK condition, 
through estimation of the KM coefficients,  
and then have checked the validity of the Pawula's theorem. 
This lets us to 
apply the reduced formula of the FP Equation~(\ref{e:wm:fop}), 
which describes evolution of the probability density function $P(b, \tau)$. 

Following our initial discussion, 
we must now verify whether the data time series under study is stationary. 
Generally, if a time series exhibits no trend, has a constant variance over time, 
and a consistent autocorrelation function over time, then it is classified as stationary. 
Such time series are also much easier to model. 
There are a variety of ways to evaluate this feature of any time series. 
One of such method is the Augmented \citeauthor{DicFul79} (\citeyear{DicFul79}) 
test. 
This test uses the following null and alternative hypotheses: $\mathbf{H_0}$: 
The time series is non-stationary, vs. $\mathbf{H_1}$: The time series is stationary. 
When the \textit{p}-value is less than $0.05$, then the null hypothesis can be rejected 
and it can be concluded that the time series is stationary.
In fact, after performing such a statistical test, 
we have determined that in cases (a) and (b), the respective \textit{p}-values are $<0.01$, 
indicating that the null hypothesis can be rejected. 
Thus, these magnetic field strength $B = |\mathbf{B}|$ time series are stationary. 
However, in case (c) where a much smaller data sample is available the \textit{p}-value is equal to $0.154$, 
hence we have failed to reject the null hypothesis. 
The result suggests that the time series is non-stationary 
and has some time-dependent structure with varying variance over time.

Once again, 
there are various methods of eliminating trends and seasonality, 
which define non-stationary time series. 
Trends can cause the mean to fluctuate over time, while seasonality can lead to changes in the variance over time.
The most straightforward approach to address this issue is the differencing technique, 
a common and frequently used data transformation that is applied for making time series data stationary. 
Differencing is achieved by subtracting the previous observation from the current one. 
Following notation in  Eq.~(\ref{e:wm:tau}),  
this  can simply be written as $b(t)=B(t)-B(t-1)$. 
To reverse this process, the prior time step's observation must be added to the difference value. 
The practice of computing the difference between successive observations is referred to as a \textsf{lag-1} difference.
The number of times that differencing is carried out is referred to as the order of differentiation. 
Fortunately, in our case (c), applying the \textsf{lag-1} (order $1$) difference operation has been sufficient 
to get rid of non-stationarity. 
The augmented Dickey-Fuller test has yielded a \textit{p}-value of less than $0.01$, 
thus the null hypothesis could be rejected, 
indicating that the analyzed $B = |\mathbf{B}|$ time series is stationary.

We have used one of the exploratory data analysis approaches called unsupervised binning method 
(compare with normalized histogram method) to make bins (histogram’s boxes)  
and to obtain the empirical conditional probability density functions 
$P(b_1,\tau_1|b_2,\tau_2)$, for $0< \tau_1 < \tau_2$ directly from the analyzed data. 
First, we have estimated the empirical joint PDF $P(b_1,\tau_1; b_2,\tau_2)$ 
by counting the number of different pairs $(b_1, b_2)$ on a 2-dimensional grid 
of equal-width data bins (small intervals). 
This \textit{bins} integer should be neither too large, 
such that each bin no longer contains a significant quantity of points, nor too small, 
such that any dependency of the drift and diffusion coefficients on the state variable cannot be detected.
Next, we have performed the normalization such that the integral over all bins is equal to $1$ 
(note that the sum will not be equal to $1$ unless bins of unity width are chosen). 
Similarly, the empirical one-dimensional PDF $P(b_2,\tau_2)$ can be estimated 
with the use of a one-dimensional grid of bins (and carrying-out the normalization), 
and the empirical conditional PDFs are obtained using Eq.~(\ref{e:wm:cond}) 
directly (in a numerical sense).

\begin{figure} 
    \centering
    \includegraphics[scale=0.48]{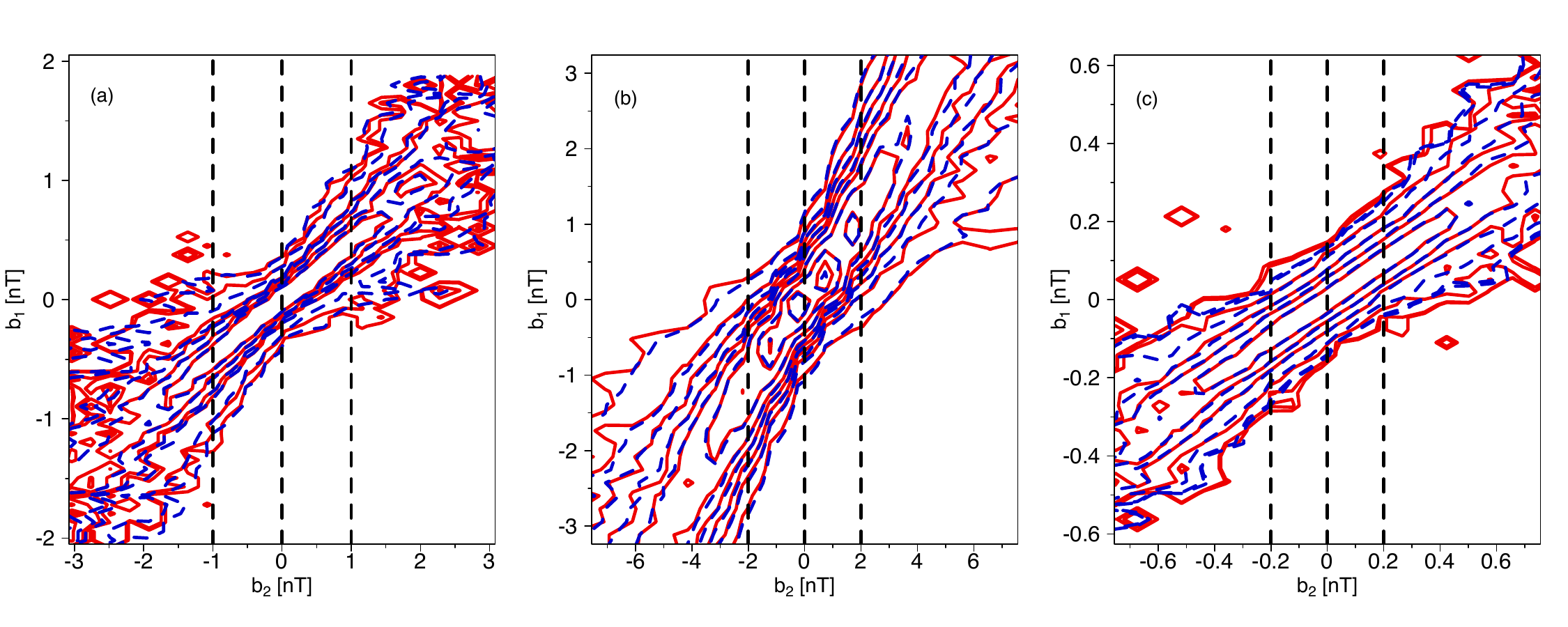}
    \caption{Comparison of observed contours (red solid curves) of conditional probabilities at various time scales $\tau$, 
    with reconstructed contours (blue dashed curves) according to the Chapman--Kolmogorov (CK) condition, 
    recovered by the use of \textit{MMS} Magnetic Field total magnitude $B = |\mathbf{B}|$ 
    in the magnetosheath: (a) just behind the bow shock (BS), (b) inside the magnetosheath (SH), and (c) near the magnetopause (MP), 
    corresponding to the spectra in Fig.~\ref{f:wm:data}.}
    \label{f:wm:conturs}
\end{figure}
 
In such a way, we have found the empirical conditional probability density functions from the analyzed data,
which are shown by red continuous contours in Fig.~\ref{f:wm:conturs}. 
They are compared here with the theoretical conditional PDFs that are solutions of the CK condition of Eq.~(\ref{e:wm:chk}) 
displayed by blue dashed contours,  
which are $2$-dimensional representation of $3$-dimensional data. 
Such a comparison is seen in Fig.~\ref{f:wm:conturs} for the magnetic field increments ${b}$, 
at the various scales: in cases (a) and (c) 
$\tau_1 = 0.02$ s, $\tau’ = \tau_1 + \Delta t_B = 0.0278$ s, $\tau_2 = \tau_1 + 2 \Delta t_B = 0.0356$~s, 
where $\Delta t_B = 0.0078$ s, and in case (b) $\tau_1 = 0.2$ s, $\tau’ = \tau_1 + \Delta t_B = 0.2625$~s, 
$\tau_2 = \tau_1 + 2 \Delta t_B = 0.325$ s, where $\Delta t_B = 0.0625$~s. 
The depicted subsequent isolines correspond to the following decreasing levels of the conditional PDFs, 
from the middle of the plots, for following magnetic field increments ${b}$: 
case (a) 2, 1.1, 0.5, 0.3, 0.05, 0.01; 
case (b) 5, 1, 0.7, 0.45, 0.3, 0.22, 0.15, 0.1, 0.05; 
and case (c) 7, 3.3, 1.3, 0.3, 0.08, 0.06. 
This is rather evident that the contour lines corresponding to these two empirical and theoretical  
probability distributions are nearly matching for all three cases. 
Thus, it appears that the CK condition of Eq.~(\ref{e:wm:chk}) is 
sufficiently well satisfied. 

\begin{figure}
    \centering
    \includegraphics[scale=0.72]{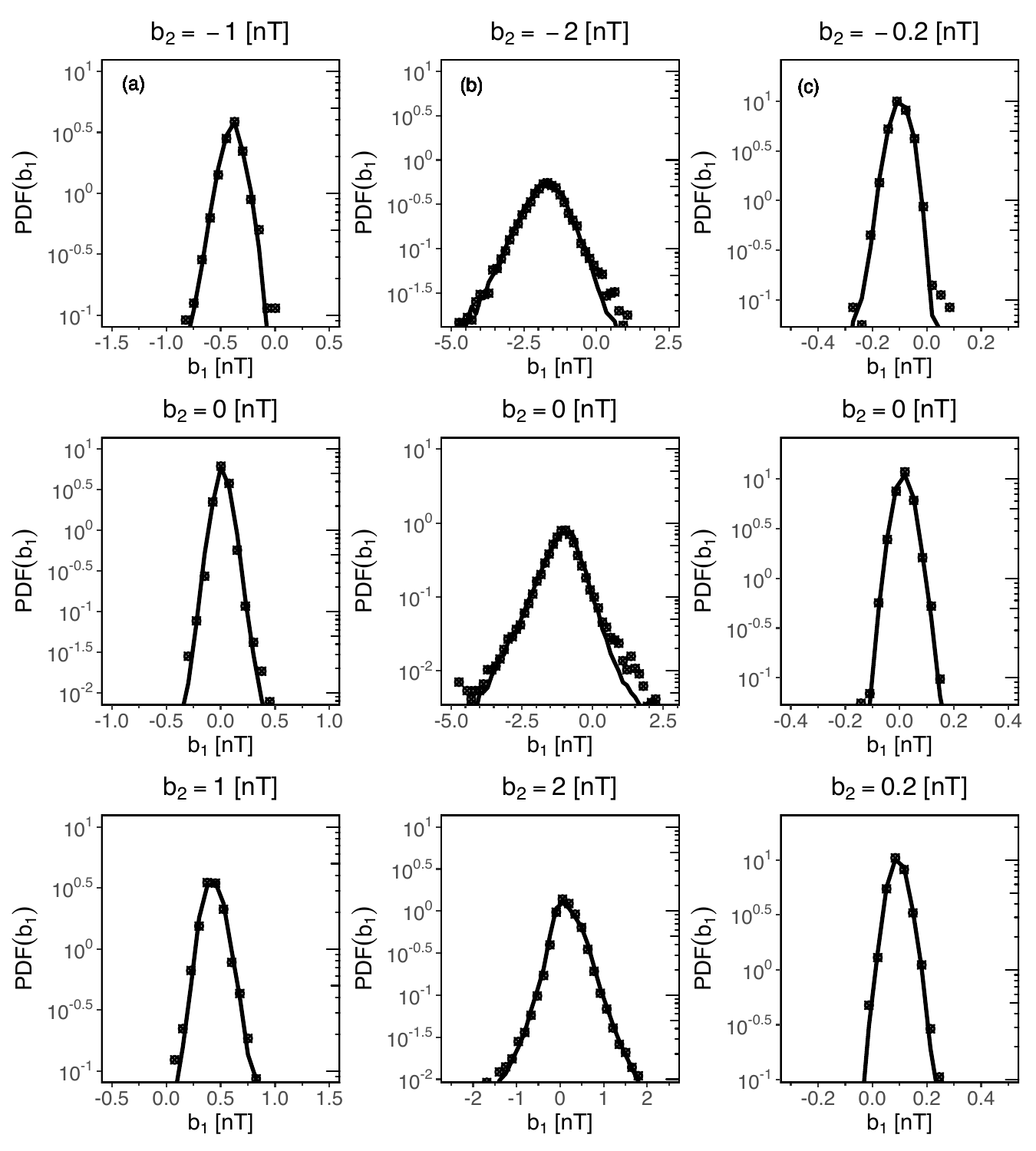}
    \caption{Comparison of cuts through $P(b_1,\tau_1|b_2, \tau_2$) for the fixed value of the strength 
    of the magnetic field total magnitude $B = |\mathbf{B}|$ in the magnetosheath: (a) just behind the bow shock (BS), (b) inside the magnetosheath (SH), and (c) near the magnetopause (MP), with increments $b_2$ with  $\tau_1$ = 0.02 s, $\tau'$ = 0.0278 s, and $\tau_2$ = 0.0356 s in cases (a) and (c), and with $\tau_1$ = 0.2 s, $\tau'$ = 0.2625 and $\tau_2$ = 0.325 s in case (b).}
    \label{f:wm:cuts}
\end{figure}

Next, in the corresponding Fig.~\ref{f:wm:cuts}, 
we have verified again the CK condition of Eq.~(\ref{e:wm:chk}). 
Intuitively speaking (and somehow informally), what we see in Fig.~\ref{f:wm:conturs} is 
just a view 'from the top' of the 3-dimensional shape, 
while in Fig.~\ref{f:wm:cuts} the orthogonal cuts are depicted. 
Again, we have compared these cuts through the conditional probability density functions 
for particular chosen values of parameter $b_2$, which can be seen at the top of each plot. 
As is evident, the cuts through the empirical probability density functions coincide rather well 
with the cuts through the theoretical probability density functions, 
providing good fits in all of the analyzed cases.  
Admittedly, only in case (b) for $b_2 = 0$ [nT] the cuts points deviate from the lines in tails, 
but it seem to be caused by the data outliers, which can eventually be further eliminated.  
It is necessary to mention that after 
such a comparison for different values of $(\tau_1, \tau'$, $\tau_2)$, 
we have found that the CK condition of Eq.~(\ref{e:wm:chk}) is satisfied for ${b}$ 
up to a scale of approximately $100 \Delta t_B = 0.78$ s in case (a), 
to about $150 \Delta t_B = 9.375$ s in case (b), 
and around $40 \Delta t_B = 0.312$ s in case (c), 
thus indicating that the 
stochastic fluctuations have Markov properties.

To verify Pawula's theorem, which states that 
if the fourth-order coefficient is equal to zero, 
then $D^{(k)}(b,\tau) = 0, \; k \ge 3$, 
it is necessary to estimate the $D^{(1)}(b,\tau)$, $D^{(2)}(b,\tau)$ and $D^{(4)}(b,\tau)$ coefficients 
using our experimental data. 
The standard procedure for calculating these values is to use an extrapolation method 
such as a piecewise linear regression to estimate the respective limits in  Eq.~(\ref{e:wm:Dk}).
However, as already mentioned in Section~\ref{sec:wm:meth}, 
the similar results are obtained by simplifying the problem of finding these coefficients, 
by using Eq.~(\ref{e:wm:cef_app}), 
which enables us to estimate these values using the adequately scaled $M^{(k)}(b,\tau,\tau')$ coefficients. 
In our situation, the time resolution $\Delta t_B$ is equal to $7.8$ ms in case (a) and (c), 
while in case (b) it is $62.5$ ms.
Thus, given the conditional probabilities $P(b_1,\tau_1|b_2,\tau_2)$, for $0< \tau_1 < \tau_2$, 
we have calculated these central moments directly from Eq.~(\ref{e:wm:Mk}),
using the obtained empirical data by counting the numbers $N(b',b)$ of occurrences of two fluctuations $b'$ and $b$. 
Given that the errors of $N(b',b)$ might be simply determined by $\sqrt{N(b',b)}$, then, in a similar way, 
it is possible to calculate the errors for the conditional moments $M^{(k)}(b,\tau,\tau')$. 
Consequently, scaling these values according to Eq.~(\ref{e:wm:cef_app}), 
we have obtained the empirical KM coefficients. 
By examination of the $M^{(k)}(b,\tau,\tau')$ and $D^{(k)}(b,\tau)$ coefficients, 
we can observe that they both exhibit the 
same dependence on $b$.

\begin{figure}
    \centering
    \includegraphics[scale=0.5]{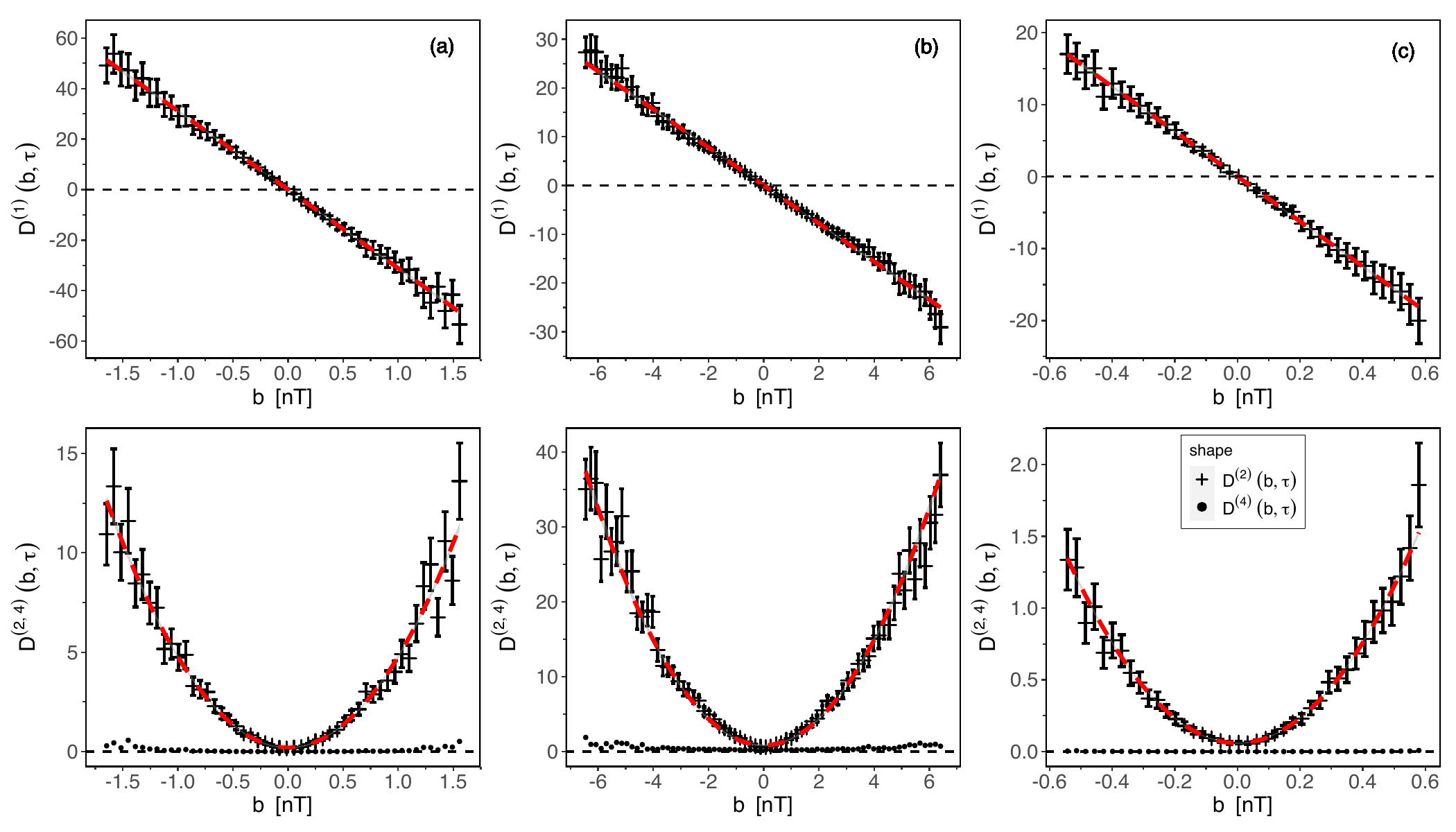}
    \caption{The first and second limited-size Kramers--Moyal coefficients 
    determined by the magnetic field increments $b$ for a total strength of magnetic field $B = |\mathbf{B}|$ 
    in the magnetosheath: (a) just behind the bow shock (BS), (b) inside the magnetosheath (SH), and (c) near the magnetopause (MP). 
    The dashed red lines show the best choice fits to the calculated values 
    of $D^{(1)}(b,\tau)$ and $D^{(2)}(b,\tau)$ with $D^{(4)}(b,\tau)$~=~0, according to the Pawula's theorem.}
    \label{f:wm:km}
\end{figure}

The results of this  analysis are shown in Fig.~\ref{f:wm:km},
where on the upper part we have depicted the first order coefficient depending on $b$, 
while at the bottom we have shown the second and fourth order coefficients depending on $b$, 
for all three cases (a), (b), and (c). 
Moreover, for each case, we have provided the calculated confidence intervals (error bars).
It is demonstrated, that the fit for $D^{(1)}(b,\tau)$ coefficient is a linear function of $b$ 
and for $D^{(2)}(b,\tau)$ is a~quadratic function of $b$, for $\Delta t_B = 0.0078$ s in cases (a) and (c), and $\Delta t_B = 0.0625$ s in case (b). 
In fact, we have checked that the same fits are reasonable up to even $150 \Delta t_B$ for all three analyzed cases. 
This  means that in this instance, there should be no difficulties with fitting the polynomials for different $\Delta t_B$.

As seen at the bottom part of Fig.~\ref{f:wm:km} of cases (a) and (c), 
it is evident that the Pawula's theorem is clearly satisfied. 
On the other hand, in case (b) it might be not so  obvious. 
For instance, for $b \approx -6.2$ nT, we can see that the value of $D^{(4)}(b, \tau)$ is somewhat greater than zero. 
In this case, we can use the somewhat weaker version of this theorem, 
which states that it is sufficient to check if $D^{(4)}(b,\tau) \ll [D^{(2)}(b,\tau)]^2$, 
for all $b$ \citep[see][]{Ris96, Rinet16}. 
Thus, in this situation, we have $[D^{(2)}(b,\tau)]^2 \approx 1225$, 
which is significantly larger than $D^{(4)}(b,\tau) \approx 1$, for $b \approx -6.2$ nT. 
Therefore, it is reasonable to conclude that 
the Pawula's theorem is sufficiently well fulfilled in all of the analyzed cases. 
Hence we can assume that the Markov process is described by the FP Equation~(\ref{e:wm:fop}).

In order to find the analytical solution of the FP Equation~(\ref{e:wm:fop}), 
we have proposed certain approximations of the lowest order KM coefficients. 
As previously discussed (see Fig.~\ref{f:wm:km}), 
it is straightforward that $D^{(1)}(b,\tau)$ exhibits a linear dependence, 
whereas $D^{(2)}(b,\tau)$ displays a quadratic dependence on $b$. 
Consequently, it is reasonable to assume the following parametrization:
\begin{equation} 
\left\{
\begin{aligned} 
    D^{(1)} (b, \tau) & = - a_1 (\tau) b, \\
    D^{(2)} (b, \tau) & = a_2 (\tau) + b_2 (\tau) b^2. 
    \label{e:wm:D12}
\end{aligned} 
\right.
\end{equation}
where the relevant parameters $a_1 >0, a_2 > 0$, and $b_2 > 0$ depend on temporal scale $\tau > 0$.  
Moreover, it appears that all of these parameters exhibit a~power-law dependence on temporal scale $\tau$:
\begin{equation}
\left\{
\begin{aligned} 
    a_1(\tau) = A \tau^{\alpha}; \\
    a_2(\tau) = B \tau^{\beta}; \\
    b_2(\tau) = C \tau^{\gamma}, 
    \label{e:wm:P-L}
\end{aligned} 
\right.
\end{equation}
where the values for all of the logarithmized parameters $A, B, C \in \mathbb{R}$, 
as well as the $\alpha, \beta, \gamma \in \mathbb{R}$ are given in Table~\ref{t:wm:tab1}. 

\begin{table} 
\centering
\caption{The fitted parameters for power-law dependence of first- and second-order Kramers–Moyal (KM) coefficients 
of Eqs.~(\ref{e:wm:D12}) and (\ref{e:wm:P-L}) as functions of scale $\tau$} 
\label{t:wm:tab1}
\resizebox{17cm}{!}{
\begin{tabular}{p{1.6cm}p{2.6cm}p{2.6cm}p{2.6cm}p{2.6cm}p{2.6cm}p{2.6cm}}
\hline
 Case & $\log_{10} (A)$ & $\alpha$ & $\log_{10} (B)$ & $\beta$ & $\log_{10} (C)$ & $\gamma$ \\
 \hline
 (a) & $0.6989 \pm 0.0225$ & $-1.1191 \pm 0.0089$ & $-0.4946 \pm 0.1259$ & $1.1631 \pm 0.0498$ & $0.5854 \pm 0.0706$ & $-1.7325 \pm 0.0279$ \\
 (b) & $0.1837 \pm 0.0139$ & $-1.0417 \pm 0.0100$ & $-0.4666 \pm 0.0160$ & $0.5425 \pm 0.0116$ & $0.4183 \pm 0.0163$ & $-1.2233 \pm 0.0118$ \\
 (c) & $0.7791 \pm 0.0079$ & $-1.1055 \pm 0.0057$ & $-0.5893 \pm 0.0126$ & $1.0002 \pm 0.0091$ & $0.5011 \pm 0.0274$ & $-1.7646 \pm 0.0199$ \\
 \hline
\end{tabular}
}
\end{table}

It is important to emphasize that the functional dependencies of the coefficients $a_1(\tau)$, $a_2(\tau)$, and $b_2(\tau)$ on $\tau$ 
given by Eq.~(\ref{e:wm:P-L}) are merely parametrizations of the empirical results. 
In fact, here power-laws have been selected, 
because they have adequately described the observed values with sufficient accuracy. 
Nevertheless, some alternative theoretical analyses may lead to 
slightly different functional dependence \cite[see][]{Renet01}.
Admittedly, it turned out that the values of the fitted parameters 
can slightly be different from those that fit exactly  
the solution of the FP Equation~(\ref{e:wm:fop}). 
\cite{Renet01} has also highlighted the asymmetry of the fit $D^{(2)}(b,\tau)$ on $b$, 
which is also present in our analysis (especially in case (c), and to a lesser degree in case (a)).

\begin{figure} 
    \centering
    \includegraphics[scale=0.6]{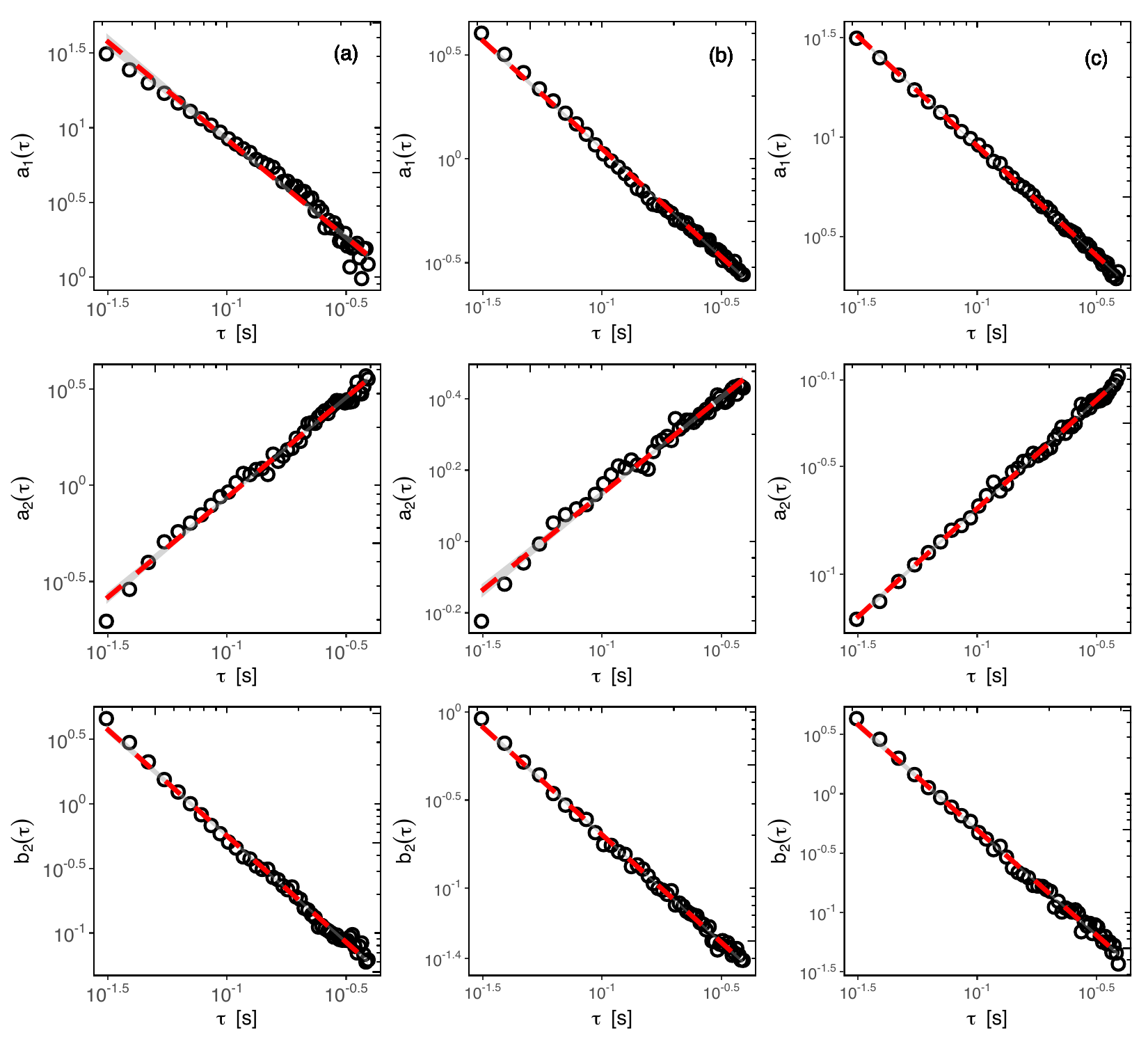}
    \caption{Linear dependence of the parameters $a_1, a_2, b_2$ (see Eq.~(\ref{e:wm:P-L})) 
    on the double logarithmic scale $\tau$ ($\log$-$\log$ plot), for the magnetic field overall intensity $B = |\mathbf{B}|$ 
    in the magnetosheath: (a) just behind the bow shock (BS), (b) inside the magnetosheath (SH), and (c) near the magnetopause (MP).
    The dashed red lines, with the standard error of the estimate illustrated by gray shade, 
    show the best choice fits to the calculated values.}
    \label{f:wm:fit}
\end{figure}

The obtained fits to the \textit{MMS} observations in the magnetosheath 
are depicted in Fig.~\ref{f:wm:fit}, for each case (a), (b), and (c), 
showing the dependence of KM coefficients parameters on scale $\tau>0$. 
Since  our data contain a~multitude of relatively low values and a few exceedingly large values, 
which would render a linear graph rather unreadable, instead of using a standard linear graph, 
we have decided to employ logarithmic scales  
for both the vertical and horizontal axes (so called: $\log-\log$ plot).
Such a procedure is rather straightforward. 
For example, for the first row of Eq.~(\ref{e:wm:P-L}), 
taking the logarithm of both sides one obtains: $\log(a_1(\tau)) = \alpha \log (\tau) + \log(A)$, 
which is a special case of a linear function, 
with the exponent $\alpha$ corresponding to the slope of the line.
The value of $\log(A)$ corresponds to the intercept of a $\log(a_1(\tau))$-axis,
while the $\log (\tau)$-axis is intercepted at $\log A/(-\alpha)$.
We have opted for this approach to enhance the clarity of the presentation. 
Therefore, since we have used both the logarithmic scales 
the respective power-laws appear as straight lines in Fig.~\ref{f:wm:fit}. 
Similarly, the graphical representations for all the parameters $a_1$, $b_1$, and $b_2$ 
of Eqs.~(\ref{e:wm:D12}) and (\ref{e:wm:P-L}), which we have provided, 
are helpful for identifying correlations and determining respective constants 
$A$, $B$, $C$ and $\alpha<0$, $\beta>0$, $\gamma<0$ 
in Table~\ref{t:wm:tab1} \citep[cf.][]{Macet23}.

After performing a careful analysis of the \textit{MMS} magnetic field magnitude ${B}$ data, 
our findings indicate that the power-law dependence is applicable for the values of: 
$\tau \lesssim 100 \Delta t_B = 0.78$ s in case (a); 
$\tau \lesssim 150 \Delta t_B = 9.375$ s in case (b); 
$\tau \lesssim  50 \Delta t_B = 0.39$ s in case (c), 
and for some larger scales, say $\tau \gtrsim \tau_G$, 
the shapes of the probability density functions appear to be closer to  Gaussian. 
However, despite the satisfactory results obtained at these small kinetic scales, 
a more intricate functional dependence (possibly polynomial fits) is characteristic for much higher scales, 
in particular, in the inertial domain \citep{StrMac08a,StrMac08b}.

As a result of our investigations, 
we are able to obtain analytical stationary solutions $p_s(x)$ given by  Eq.~(\ref{e:wm:sta}) 
following from the FP Equation~(\ref{e:wm:fop}), 
which can be expressed by a continuous kappa distribution 
(also known as Pearson's type VII distribution), 
which exhibits a deviation from the normal Gaussian distribution. 
The probability density function (PDF) of this distribution is of a~given form:
\begin{equation} \label{e:wm:kap}
p_s(b) = \frac{N_o}{\Big[1 + \frac{1}{\kappa} \Big(\frac{b}{b_o}\Big)^2\Big]^{\kappa}}, 
\end{equation}
where, for $a_2(\tau) \neq 0, \: b_0(\tau) \neq 0$, 
we have a shape parameter  $\kappa = 1 + a_1(\tau) / [2 b_2(\tau)]$ and $b_0^2 = a_2(\tau) / [b_2(\tau) + a_1(\tau)/2]$, 
while $N_0 = p_s(0)$ satisfies the normalization $\int_{- \infty}^{\infty} p_s(b) db = 1$. 
By substituting $p_s(b)$ to this integral we find that:
\begin{equation} \label{e:wm:norm}
    N_0 = \frac{\Gamma(\kappa)}{\Gamma(\kappa- \frac{1}{2}) b_0 \sqrt{\pi \kappa}},
\end{equation}
where, this time, $\Gamma(\kappa) = \int_0^{\infty} b^{\kappa - 1} \mathrm{e}^{-b} db$, $\text{Re}(\kappa) > 0$ 
is a mathematical gamma function (Euler integral of the second kind), 
as defined for all complex numbers with a~positive real part.

It is worth noting that kappa distribution, as represented by Eq.~(\ref{e:wm:kap}), 
approaches the normal Gaussian (Maxwellian) distribution for large values of scaling parameter $\kappa$. 
To be precise, as $\kappa \to \infty$, the following well-known formula can approximately be satisfied:
\begin{equation}
    \lim_{\kappa \to \infty} p_s (b) = N_0 \mathrm{\exp}\Big(-\frac{b^2}{2\sigma^2}\Big)
\end{equation}
with the scaling parameter $b_0$ related to the standard deviation $\sigma = b_0/\sqrt{2}$.
This time the parameter $N_0 = p_s(0)$ satisfies the elementary normalization $N_0 = \frac{1}{\sigma \sqrt{2 \pi}}$. 

The numerical results of fitting the empirical \textit{MMS} data to the given distributions 
and determining the relevant parameters of Eq.~(\ref{e:wm:kap}) are as follows:  
$\kappa = 1.5179$, $b_0 = 1.9745$, and $N_0 = 0.68438$ for $B$ in case (a); 
$\kappa = 1.3758$, $b_0 = 2.6955$, and $N_0 = 0.34375$ in case (b); 
with $\kappa = 3.5215$, $b_0 = 1.7313$, and $N_0 = 1.1866$ in case (c).  
These values of $\kappa$ would correspond to the nonextensivity parameter 
of the generalized (Tsallis) entropy $q = 1 - 1/\kappa$
\citep[e.g.,][]{BurVin05}. 
In our case this is given by $q = \frac{a_1(\tau)}{a_1(\tau) + 2 b_2(\tau)}$  
and is equal to $0.341$ in case (a), $0.273$  in case (b), and $0.716$ in case (c). 
The extracted values of the $\kappa$ and $q$ parameters provide 
robust measures of the departure of the system from equilibrium.
We see that these values are similar to  $q \sim 0.5$ for $\kappa \sim 2$
reported for the Parker Solar Probe (\textit{PSP}) data by \cite{Benet22}. 

Now, by using the system of Eqs.~(\ref{e:wm:D12}) with Eq.~(\ref{e:wm:fop}), 
we have arrived at the following formula \citep{Macet23}:
\begin{eqnarray}
    \Big(a_2(\tau) + b_2(\tau) b^2 \Big) \frac{\partial^2 P(b, \tau)}{\partial b^2} 
    + \Big( a_1(\tau) + 4 b_2(\tau) \Big)  b \frac{\partial P(b, \tau)}{\partial b} + & {} & \nonumber\\
    + \frac{\partial P(b, \tau)}{\partial \tau} + \Big(a_1(\tau) + 2 b_2(\tau) \Big) P(b, \tau) = 0. 
    \label{e:wm:fop3}
\end{eqnarray}
This implies that the FP  Equations~(\ref{e:wm:fop2}) and (\ref{e:wm:fop3}) 
are expressed in terms of a second-order parabolic partial differential equation. 
Thus, through the implementation of the numerical Euler integration scheme, 
which has been verified for stationary solution $\frac{\partial P(b, \tau)}{\partial \tau} = 0$, 
we are able to successfully solve the non-stationary FP equation numerically. 
Our results are in line with those obtained by~\cite{Rinet16} using 
the statistical modeling package in programming language \textsc{R}.

\begin{figure}%
    \centering
    \includegraphics[scale=0.49]{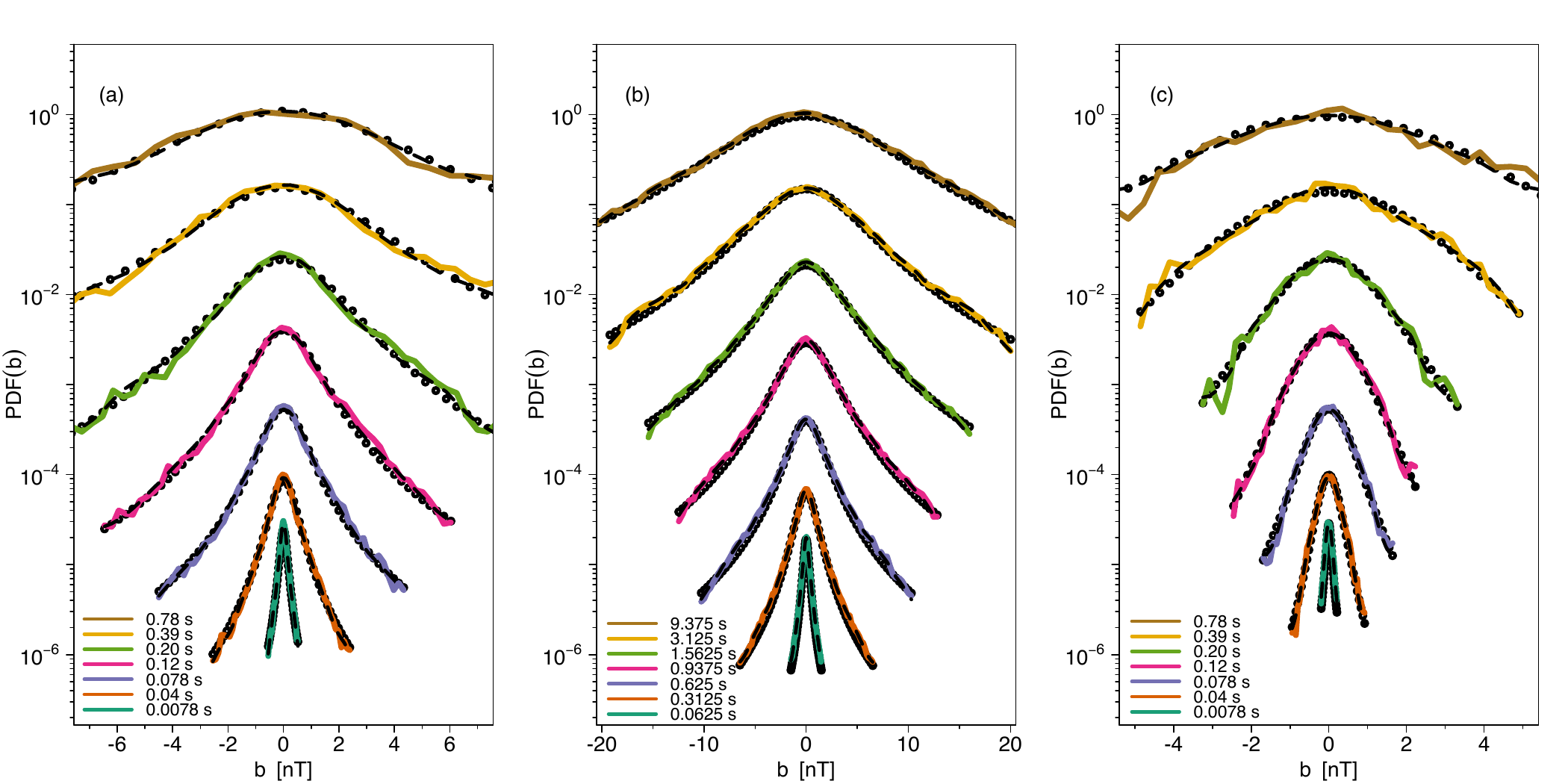}
    \caption{The empirical probability density functions (various continuous colored lines) 
    for a total strength of magnetic field $B = |\mathbf{B}|$, which correspond to spectra in Fig.~\ref{f:wm:data}, 
    compared with the non-stationary (dashed lines) and the stationary (open circles) solutions of the FP equation, 
    for various time scales (shifted from bottom to top) $\tau$ = 0.0078, 0.04, 0.078, 0.12, 0.2, 0.39, and 0.78 s in cases (a) and (c), 
    and $\tau$ = 0.0625, 0.3125, 0.625, 0.9375, 1.5625, 3.125, 9.375 s in case (b).}
    \label{f:wm:fokker}
\end{figure}

Figure~\ref{f:wm:fokker} shows the findings resulting from our analysis based on the \textit{MMS} data. 
Here we compare the solutions of the FP Equation~(\ref{e:wm:fop}) 
with the empirical probability density functions of $P(b,\tau)$: 
(a) near the bow shock (BS), (b) inside the magnetosheath (SH), 
and (c) near the magnetopause (MP) for various scales $\tau$ 
(not greater than $\tau_{G}$). 
The displayed plotted curves, in each case, are as follows: 
the stationary solution (denoted by open circles), the non-stationary solutions (marked with dashed lines), 
and the empirical PDFs (indicated with various colored continuous lines). 

Further, in cases (a) and (c) the corresponding time scales are: 
$\tau = 0.0078, 0.04, 0.078, 0.12, 0.2, 0.39$, and $0.78$ s; 
whereas in case (b) these scales are equal: $\tau = 0.0625, 0.3125, 0.625, 0.9375, 1.5625, 3.125$, and $9.375$ s.
The corresponding curves are shifted in the vertical direction from bottom to top 
for even better clarity of presentation. 
It is also worth noting that we have used the semi-logarithmic scale $\tau$,
what is useful when dealing with data that covers a broad range of values. 
On this scale, the vertical scale is logarithmic (base $10$) axis, 
which means that the separation between the ticks on the graph is proportional to the logarithm of PDF, 
while the horizontal $b$-axis is a standard linear scale, and the ticks are evenly spaced. 

What is important to note from this picture are the peaked leptokurtic shapes of PDFs 
and corresponding stationary solutions. 
Namely, in case (a) the peak (with large kurtosis) is present for scales up to 
$\thicksim0.5$ s; in case (b) up to about $\thicksim3$ s; and in case (c) up to $\thicksim0.25$ s. 
For these levels selected for each case the PDF becomes closer to  Gaussian 
(i.e., approximately parabolic shape on the graph with the semi-logarithmic scales),  
as expected for large values of $\kappa$.
In case (c) we can see more jumps in fluctuations, i.e., the curves are not so smooth. 
Fluctuations are quite evident in both the empirical curves and the theoretical solutions, 
so it seems that some numerical noise is present in the tails of the PDFs. 
Admittedly, reducing noise is a tricky issue, although the easiest way is to artificially smooth using the simple moving average. 
Therefore, we have tried this procedure for $n = 1, 2, 3$ steps and it has appeared that the $n=3$ choice is 
sufficient.

\begin{figure}%
    \centering
    \includegraphics[scale=0.49]{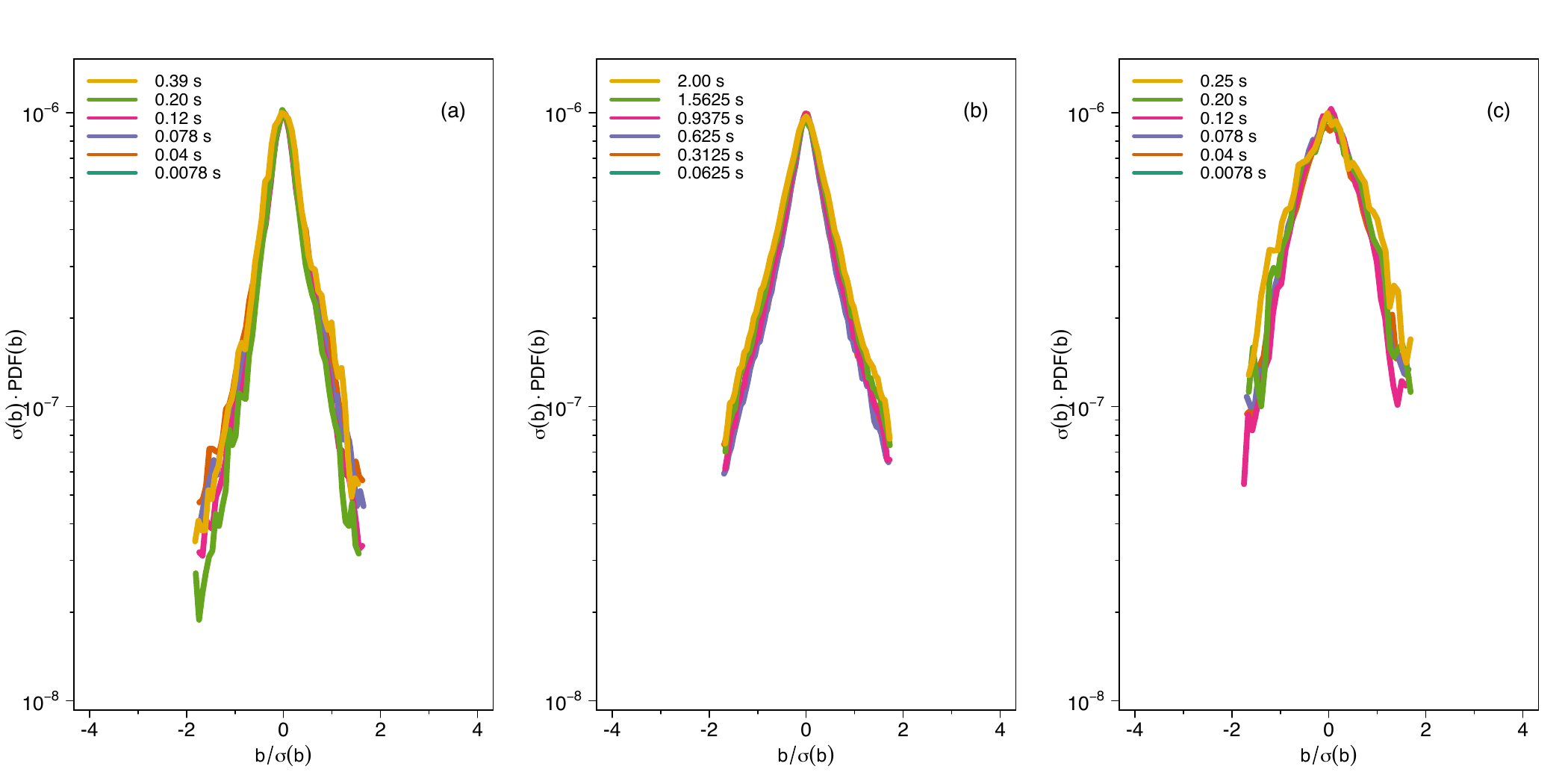}
    \caption{%
    A collapse of probability density functions of $b_\tau$ 
    (compare Fig.~\ref{f:wm:fokker}), which are scaled by the corresponding standard deviations 
    (see Eqs.~(\ref{e:wm:scl1}) and~(\ref{e:wm:scl2})), 
    for the small time scales $\tau$ stopping at approximately: $\tau \sim 0.4$ s. 
    in case (a), $\tau \sim 2.0$ s in case (b), and $\tau \sim0.25$ s in case (c).}
    \label{f:wm:collaps}
\end{figure}

Figure~\ref{f:wm:collaps} depicts finally the probability density functions (PDFs) 
of fluctuations of the strength of the magnetic field $b_{\tau}$ 
rescaled by the standard deviations $\sigma(b_{\tau})$ in the following way:
\begin{align}
b_{\tau} & \longrightarrow \frac{b_{\tau}}{\sigma (b_{\tau})}, \label{e:wm:scl1}\\
\text{PDF}(b_{\tau}) &\longrightarrow \sigma (b_{\tau}) \cdot \text{PDF}(b_{\tau}). \label{e:wm:scl2}
\end{align}
In this way, we can define a master curve for the shape of the PDFs. 
Again, we have used the logarithmic scale on the vertical axis. 
We also see that the rescaled curves are consistent with the stationary solutions  
of Eq.~(\ref{e:wm:kap}), as marked with open circles in Fig.~\ref{f:wm:fokker}. 
It should be noted that all the curves in Fig.~\ref{f:wm:collaps} are very close to each other for small scales. 
However, for larger $\tau = 50$ or $100 \Delta t_B$ these shapes deviate from the master curve 
and naturally tend to the well known Gaussian shape. 
We see that the shape of the PDFs obtained from the \textit{MMS} data 
exhibits a global \textit{scale-invariance} in the magnetosheath up to scales of $\sim2$~s. 
A similar collapse has also been reported with the \textit{PSP} data at subproton scales \citep{Benet22}. 
Figure \ref{f:wm:collaps} shows that fluctuations in the magnetosheath are described by a stochastic process.
Admittedly, the mechanism of generation of these magnetic fluctuation at small kinetic scale is not known, 
but the this results suggest some universal characteristics of the  processes.
An alternative point of view has recently been proposed by \cite{Caret22}.  

\section{Conclusions}
\label{sec:wm:con}


Following our studies in the space plasmas at large inertial scales \citep{StrMac08a,StrMac08b},  
we have examined time series of the strength of magnetic fields 
in different regions of the Earth's magnetosheath, 
where the spectrum steepens at subproton scales \citep{Macet18}.
With the highest resolution available on the \textit{MMS}, 
the data samples just after the bow shock and near the magnetopause are stationary 
and for somewhat lower resolution deep inside the magnetosheath 
the deviations from stationarity are small and could well be eliminated. 
Basically, in all these cases the stochastic fluctuations exhibits Markovian features. 
We have verified that the necessary Chapman--Kolmogorov condition is well satisfied,
and the probability density functions are consistent with the solutions of this condition.

In addition, the Pawula's theorem is also well satisfied resulting in 
the Fokker--Planck equation reduced to drift and diffusion terms. 
Hence, this corresponds to the generalization of Ornstein--Uhlenbeck process. 
Further, the lowest Kramers--Moyal coefficients have linear and quadratic dependence 
as functions of the magnetic field increments.
In this way, the power-law distributions are well recovered throughout the entire magnetosheath. 
For some moderate scales we have the kappa distributions 
described by various peaked shapes with heavy tails. 
In particular, for large values of the kappa parameter 
these distributions are reduced to the normal Gaussian distribution. 

Similarly as for the \textit{PSP} data, 
the probability density functions of the magnetic fields measured onboard the \textit{MMS} 
rescaled by the respective standard deviations exhibit a universal global \textit{scale-invariance}
on kinetic scales, which is consistent with the stationary solution of the Fokker--Planck equation.
We hope that all these results, especially those reported at small scales, 
are important for a better understanding of the physical mechanism 
governing turbulent systems in space and laboratory.

\section*{Acknowledgements}


We thank Marek Strumik for discussion on the theory of Markov processes. 
We are grateful for the efforts of the entire \textit{MMS} mission, 
including development, science operations,and the Science Data Center at the University of Colorado. 
We benefited from the efforts of  T.~E. Moore as Project Scientist, C.~T. Russell and the magnetometer team. 
We acknowledge B. L. Giles, Project Scientist for information about the magnetic field instrument,
and also to D. G. Sibeck and M. V. D. Silveira for discussions 
during previous visits by W. M. M to the NASA Goddard Space Flight Center.

This work has been supported by the National Science Centre, Poland (NCN), through grant
No. 2021/41/B/ST10/00823.

\section*{Data Availability}
 

The data supporting the results in this article are available through
the MMS Science Data Center at the Laboratory for Atmospheric
and Space Physics (LASP), University of Colorado, Boulder: 
\url{https://lasp.colorado.edu/mms/sdc/public/}.
The magnetic field data from the magnetometer 
are available online from \url{http://cdaweb.gsfc.nasa.gov}.
The data have been processed using the statistical programming language \textsc{R}.

\vspace{1cm}
ORCID iDs

\noindent
W. M. Macek   \url{https://orcid.org/0000-0002-8190-4620}\\ 
              \url{http://www.cbk.waw.pl/~macek}\\
D. W\'{o}jcik \url{https://orcid.org/0000-0002-2658-6068}\\



\bibliographystyle{mnras}
%



\bsp	
\label{lastpage}
\end{document}